\documentclass[a4paper,12pt]{article}
\usepackage{jheppub}
\usepackage{epsfig}
\usepackage{adjustbox,lipsum}

\definecolor{bl}{cmyk}{1,0.9,0,0.3}
\definecolor{br}{cmyk}{0,0.9,0,0.7}

\title{Sphalerons and the Electroweak Phase Transition in Models with Higher Scalar Representations}
\author[a,b,c]{Amine Ahriche}\author[d]{Talal Ahmed Chowdhury}\author[e]{and Salah Nasri}
\affiliation[a]{Department of Physics, University of Jijel, PB 98
Ouled Aissa, DZ-18000 Jijel, Algeria} \affiliation[b]{Fakult\"at
f\"ur Physik, Universit\"at Bielefeld, 33501 Bielefeld, Germany}
\affiliation[c]{The Abdus Salam International Centre for Theoretical
Physics, Strada Costiera 11, I-34014, Trieste, Italy}
\affiliation[d]{SISSA, Via Bonomea 265, 34136, Trieste, Italy}
\affiliation[e]{Department of Physics, UAE University, P.O. Box
17551, Al-Ain, United Arab Emirates} \emailAdd{aahriche@ictp.it,
chowdhu@sissa.it, snasri@uaeu.ae.ac}

\abstract{In this work we investigate the sphaleron solution in a
$SU(2)\times U(1)_X$ gauge theory, which also encompasses the
Standard Model, with higher scalar representation(s)
($J^{(i)},X^{(i)}$). We show that the field profiles describing the
sphaleron in higher scalar multiplet, have similar trends like the
doublet case with respect to the radial distance. We compute the
sphaleron energy and find that it scales linearly with the vacuum
expectation value of the scalar field and its slope depends on the
representation. We also investigate the effect of $U(1)$ gauge field
and find that it is small for the physical value of the mixing
angle, $\theta_{W}$ and resembles the case for the doublet. For
higher representations, we show that the criterion for strong first
order phase transition, $v_{c}/T_{c}>\eta$, is relaxed with respect
to the doublet case, i.e. $\eta<1$.} \keywords{sphalerons, scalar
multiplets.}

\begin{document}

\maketitle

\section{Introduction}

\label{section1}

In the Standard Model (SM), the anomalous baryonic and leptonic currents lead
to fermion number non-conservation due to the instanton induced transitions
between topologically distinct vacua of $SU(2)$ gauge fields
\cite{'tHooft:1976up,'tHooft:1976fv} and at zero temperature, the rate is of
the order, $e^{-2\pi/\alpha_{w}}$, $\alpha_{w}\sim1/30$, which is irrelevant
for any physical phenomena. However, there exists a static unstable solution
of the field equations, known as sphaleron \cite{Dashen:1974ck, Manton:1983nd,
Klinkhamer:1984di, Soni:1980ps}, that represents the top of the energy barrier
between two distinct vacua and at finite temperature, because of thermal
fluctuations of fields, fermion number violating vacuum to vacuum transitions
can occur which are only suppressed by a Boltzmann factor, containing the
height of the barrier at the given temperature, i.e. the energy of the
sphaleron \cite{Kuzmin:1985mm}. Such baryon number violation induced by the
sphaleron is one of the essential ingredients of Electroweak Baryogenesis
\cite{Shaposhnikov:1987tw, Shaposhnikov:1987pf, Arnold:1987mh,
Khlebnikov:1988sr, Dine:1989kt, Dine:1991ck} and therefore it has been
extensively studied not only in the SM \cite{Akiba:1988ay, Akiba:1989xu,
Yaffe:1989ms, Arnold:1987zg, Carson:1990jm, Klinkhamer:1990fi,
Kleihaus:1991ks, Kunz:1992uh, Brihaye:1992jk, Braibant:1993is, Brihaye:1993ud}
and but also in extended SM variants such as, SM with a singlet
\cite{Choi:1994mf, Ahriche:2007jp}, two Higgs doublet model
\cite{Kastening:1991nw}, Minimal Supersymmetric Standard Model
\cite{Moreno:1996zm}, the next-to-Minimal Supersymmetric Standard Model
\cite{Funakubo:2005bu} and 5-dimensional model \cite{Ahriche:2009yy}.

As many SM extensions involve non-minimal scalar sectors, it is instructive to
determine the behavior of the sphaleron for general $SU(2)$ scalar
representations. Although, apart from some exceptions like Georgi-Machacek
\cite{Georgi:1985nv} and isospin-3 models \cite{Kanemura:2013mc}, large Higgs
multiplets other than the doublet are stringently constrained by electroweak
precision observables. In addition, the presence of scalar multiplets with
isospin $J\geq5$ brings down the Landau pole of the gauge coupling to about
$\Lambda_{\text{landau}}\leq10$ TeV \cite{AbdusSalam:2013eya}. Moreover as
shown in \cite{Hally:2012pu, Earl:2013jsa}, by saturating unitarity bound on
zeroth order partial wave amplitude for the $2\rightarrow2$ scattering of
scalar pair annihilations into electroweak gauge bosons, one can set complex
$SU(2)$ multiplet to have isospin $J\leq7/2$ and real multiplet to have
$J\leq4$. Therefore it can be seen that large scalar representations of SM
gauge group are generally disfavored.

Still, motivated by the dark matter content and baryon asymmetry of the
universe, one can assume a hidden or dark sector with its own gauge
interactions. If the interaction between SM and hidden sector is feeble in
nature, they may not equilibrate in the whole course of the universe.
Therefore, the hidden sector can be fairly unconstrained apart from its total
degrees of freedom such that the sector doesn't change the total energy
density of the universe in such way that the universe had a modified expansion
rate in earlier times, specially at the BBN and CMB era. With this possibility
in mind, we can consider the hidden sector to have SM-like gauge structure
that contains scalar multiplets larger than doublet and also has its own
spontaneous symmetry breaking scale (the possibility of non-abelian gauge
structure in dark sector and non-SM sphaleron in symmetric phase for such
models are also addressed in \cite{Blennow:2010qp, Barr:2013tea}). For this
reason, it is interesting to ask what could be the nature of the sphaleron in
such SM-like $SU(2)\times U(1)_{X}$ gauge group with general scalar
multiplets. Furthermore, as sphaleron is linked with nontrivial vacuum
structure of non-abelian gauge theory, it is relevant to see the effect of
large scalar multiplets in hot gauge theories.

This paper is organized as follows. In section \ref{section2} we
discuss the spherically symmetric ansatz for larger scalar
multiplets and consequently calculated the energy functional and
variational equations for scalar multiplet $(J,X)$, give different
numerical results. In section \ref{section3} we investigate the
effect of $U(1)_{X}$ field on sphaleron energy and study the
sphaleron energy dependence on the scalar vev. Section
\ref{section4} is devoted to the conditions of the sphaleron
decoupling during the electroweak phase transition, and in section
\ref{section5} we conclude. In Appendix \ref{section6}, we have
presented the asymptotic solutions and their dependence on the
representation $(J,X)$.

\section{Sphalerons in General Scalar Representation}\label{section2}

\subsection{Spherically symmetric Ansatz}

The standard way to find sphaleron solution in the Yang-Mills-Higgs theory is
to construct non-contractible loops in field space \cite{Klinkhamer:1984di}.
As the sphaleron is a saddle point solution of the configuration space, it is
really hard to find them by solving the full set of equations of motion.
Instead one starts from an ansatz depending on a parameter $\mu$ that
characterizes the non-contractible loop in the configuration space and
corresponds to the vacuum for $\mu=0$ and $\pi$ while $\mu=\frac{\pi}{2}$
corresponds the highest energy configuration, in other words, the sphaleron.

Consider the scalar multiplet $Q$, charged under $SU(2)\times
U(1)_{X}$ group, is in $J$ representation and has $U(1)_{X}$ charge
$X$. Here $SU(2)$ and $U(1)_{X}$ can be applicable for both standard
model gauge group or SM-like gauge group of the hidden sector. The
generators in this representation are denoted as $J^{a}$ such that,
$Tr[J^{a}J^{b}]=D(R)\delta^{ab}$ where $D(R)$ is the Dynkin index
for the representation. As our focus is on the SM, we define the
charge operator, $\hat{Q}_{c}=J_{3}+X$ and require the neutral
component ($J_{3}=-X$) of the multiplet to have the vacuum
expectation value (vev).

The gauge-scalar sector of the Lagrangian is
\begin{equation}
\mathcal{L}=-\frac{1}{4}F^{a\mu\nu}F_{\mu\nu}^{a}-\frac{1}{4}f^{\mu\nu}
f_{\mu\nu}+(D_{\mu}Q)^{\dagger}D^{\mu}Q-V(Q),
\end{equation}
with scalar potential%
\begin{equation}
V(Q)=-\mu^{2}_{Q}Q^{\dagger}Q+\lambda_{1}(Q^{\dagger}Q)^{2}+\lambda_{2}(Q^{\dagger}J^{a}Q)^{2}.
\end{equation}
It was shown in \cite{Ahriche:2007jp} that the kinetic term of the scalar
field makes larger contribution to the sphaleron energy than the potential
term. Therefore, for simplicity, we have considered CP-invariant scalar
potential involving single scalar representation to determine the sphaleron
solution. It is straightforward to generalize the calculation for the
potential with multiple scalar fields\footnote{In fact, in the SM, one needs
large couplings between Higgs and extra scalars to trigger a strong first
order phase transition.}.

Also for convenience we elaborate,
\begin{align}
F_{\mu\nu}^{a} & =\partial_{\mu}A_{\nu}^{a}-\partial_{\nu}A_{\mu}%
^{a}+g\epsilon^{abc}A_{\mu}^{b}A_{\nu}^{c},\nonumber\\
f_{\mu\nu} & =\partial_{\mu}a_{\nu}-\partial_{\nu}a_{\mu},\nonumber\\
D_{\mu}Q &
=\partial_{\mu}Q-igA_{\mu}^{a}J^{a}Q-ig^{\prime}a_{\mu}XQ,
\label{covder}%
\end{align}
where, $g$ and $g^{\prime}$ are the $SU(2)$ and $U(1)_{X}$ gauge couplings.
The mixing angle $\theta_{W}$ is $\tan\theta_{W}=g^{\prime}/g$.

The scalar sector plays an essential role in constructing sphaleron and the
symmetry features of the ansatz partly depends on the $SU(2)$ representation
and $U(1)_{X}$ charge assignment of the scalar that acquires a vev. The
simplest possibility is to consider a spherically symmetric ansatz because
spherical symmetry enables one to calculate the solution and the energy of the
sphaleron without resorting into full partial differential equations.
Therefore one may ask, which scalar representation immediately allows the
spherical symmetric ansatz.

As pointed out in \cite{Yaffe:1989ms}, spherically symmetric configurations
are those for which an $O(3)$ rotation of spatial directions are compensated
by the combination of $SU(2)$ gauge and $SU(2)$ global transformation. The
existence of this $SU(2)$ global symmetry is manifest for the Higgs doublet as
the potential for the doublet has $SO(4)\sim SU(2)\times SU(2)$ global
symmetry which is broken by the scalar vev to $SU(2)\sim SO(3)$ symmetry that
leads to the mass degeneracy of three gauge bosons of $SU(2)$. One can
immediately see that this degeneracy will be lifted when the $U(1)_{X}$ is
turned on. Following the same reasoning, one can find other scalar multiplets
that will lead to mass degeneracy of $A_{\mu}^{a}$'s in $SU(2)$ gauge theory
after the symmetry is broken.

In the case of many scalar representations $Q^{(i)}$ with $J^{(i)}$ and charge
$X^{(i)}$, the corresponding vev's are $\langle Q^{(i)}\rangle=\frac{v_{i}%
}{\sqrt{2}}(0,..,1,..,0)^{T}$, where the non-zero neutral component quantum
numbers are $(J^{(i)},J_{3}^{(i)}=-X^{(i)})$. Now from the scalar kinetic term,%

\begin{align}
\mathcal{L} & \supset\tfrac{1}{2}g^{2}{\sum\limits_{i}}\langle
Q^{(i)\dagger}\rangle J_{a}^{(i)}J_{b}^{(i)}\langle Q^{(i)}\rangle A_{\mu}%
^{a}A^{\mu b}\nonumber\\
& =\tfrac{1}{2}g^{2}{\sum\limits_{i}}v_{i}^{2}(J^{(i)}(J^{(i)}+1)-X^{(i)2}%
)A_{\mu}^{+}A^{\mu-}+\tfrac{1}{2}g^{2}{\sum\limits_{i}}v_{i}^{2}X^{(i)2}%
A_{\mu}^{3}A^{\mu3}. \label{L}%
\end{align}
where $A^{\pm}_{\mu}=A^{1}_{\mu}\mp i A^{2}_{\mu}$. So the condition for
having equal coupling of three gauge fields to the neutral component leads to
the tree-level condition%
\begin{equation}
\rho=\frac{{\sum\limits_{i}}v_{i}^{2}(J^{(i)}(J^{(i)}+1)-X^{(i)2})}%
{2{\sum\limits_{i}}v_{i}^{2}X^{(i)2}}=1. \label{rho}%
\end{equation}
In the case of one scalar multiplet, this can be reduced to $J(J+1)=3X^{2}$.
The multiplets satisfying the above condition are $(J,X)=(\frac{1}{2},\frac
{1}{2}),(3,2)..$. Intuitively, one can consider that the scalar multiplet
enables the three gauge fields to scale uniformly like a sphere in a three
dimensional space.

\subsection{The Energy Functional and Variational Equations}

In the following we will address the energy functional and the variational
equations of the sphaleron. The classical finite energy configuration are
considered in a gauge where the time component of the gauge fields are set to
zero. Therefore the classical energy functional over the configuration is%

\begin{equation}
E(A_{i}^{a},a_{i},Q)=\int d^{3}x\left[ \frac{1}{4}F_{ij}^{a}F_{ij}^{a}%
+\frac{1}{4}f_{ij}f_{ij}+(D_{i}Q)^{\dagger}(D_{i}Q)+V(Q)\right] .
\label{enfunc}%
\end{equation}
The non-contractible loop (NCL) in configuration space is defined as map
$S^{1}\times S^{2}\sim S^{3}$ into $SU(2)\sim S^{3}$ using the following
matrix $U^{\infty}\in SU(2)$ \cite{Klinkhamer:1990fi},
\begin{align}
U^{\infty}(\mu,\theta,\phi) & =(\cos^{2}\mu+\sin^{2}\mu\cos\theta
)I_{2}+i\sin2\mu(1-\cos\theta)\tau^{3}\nonumber\label{Unif}\\
& +2i\sin\mu\sin\theta(\sin\phi\tau^{1}+\cos\phi\tau^{2}),
\end{align}
where $\mu$ is the parameter of the NCL and $\theta$, $\phi$ are the
coordinates of the sphere at infinity. Also, $\tau^{a}$ are the $SU(2)$
generators in the fundamental representation. We also define the following 1-form%

\begin{equation}
i(U^{\infty-1})dU^{\infty}=\sum_{a}F_{a}\tau^{a},
\end{equation}
which gives
\begin{align}
F_{1} & =-[2\sin^{2}\mu\cos(\mu-\phi)-\sin2\mu\cos\theta\sin(\mu
-\phi)]d\theta\nonumber\\
& -[\sin2\mu\cos(\mu-\phi)\sin\theta+\sin^{2}\mu\sin2\theta\sin(\mu
-\phi)]d\phi,\nonumber\\
F_{2} & =-[2\sin^{2}\mu\sin(\mu-\phi)+\sin2\mu\cos\theta\cos(\mu
-\phi)]d\theta\nonumber\\
& +[\sin^{2}\mu\sin2\theta\cos(\mu-\phi)-\sin2\mu\sin\theta\sin(\mu
-\phi)]d\phi,\nonumber\\
F_{3} & =-\sin2\mu\sin\theta d\theta+2\sin^{2}\theta\sin^{2}\mu
d\phi.
\end{align}

As shown in \cite{Klinkhamer:1990fi}, the NCL starts and ends at the vacuum
and consists of three phases such that in first phase $\mu\in\lbrack-\frac
{\pi}{2},0]$ it excites the scalar configuration, in the second phase $\mu
\in\lbrack0,\pi]$ it builds up and destroys the gauge configuration and in the
third phase $\mu\in\lbrack\pi,\frac{3\pi}{2}]$ it destroys the scalar configuration.

The field configurations in the first and third phases, $\mu\in\lbrack
-\frac{\pi}{2},0]$ and $\mu\in\lbrack\pi,\frac{3\pi}{2}]$ are
\begin{equation}
gA_{i}^{a}\tau^{a}dx^{i}=g^{\prime}a_{i}dx^{i}=0,
\end{equation}
and
\begin{equation}
Q=\frac{v(\sin^{2}\mu+h(\xi)\cos^{2}\mu)}{\sqrt{2}}\left(
\begin{array}
[c]{ccccc}%
0 & .. & 1 & .. & 0
\end{array}
\right) ^{T}, \label{higgs1}%
\end{equation}
with $\xi=g\Omega r$\ is radial dimensionless coordinate and $\Omega$\ is the
mass parameter used to scale $r^{-1}$, which we choose in what follows as
$\Omega=m_{W}/g$. In the second phase $\mu\in\lbrack0,\pi]$, the field
configurations are%

\begin{equation}%
\begin{array}
[b]{l}%
gA_{i}^{a}\tau^{a}dx^{i}=(1-f(\xi))(F_{1}\tau^{1}+F_{2}\tau^{2})+(1-f_{3}%
(\xi))F_{3}\tau^{3},\\
g^{\prime}a_{i}dx^{i}=(1-f_{0}(\xi))F_{3},
\end{array}
\label{ulaan}%
\end{equation}
and%
\begin{equation}
Q=\frac{v h(\xi)}{\sqrt{2}}\left(
\begin{array}
[c]{ccccc}%
0 & .. & 1 & .. & 0
\end{array}
\right) ^{T}. \label{higgs2}%
\end{equation}
Here, $f(\xi)$, $f_{3}(\xi)$, $f_{0}(\xi)$ and $h(\xi)$ are the radial profile
functions. From Eq.(\ref{ulaan}), one can see that in the spherical coordinate
system, for the chosen ansatz, the gauge fixing has led to, $A_{r}^{a}%
=a_{r}=a_{\theta}=0$. Moreover, similar to Eq.(\ref{ulaan}), the gauge fields
acting on the scalar field $Q$ can be written as
\begin{equation}
gA_{i}^{a}J^{a}dx^{i}=(1-f)(F_{1}J^{1}+F_{2}J^{2})+(1-f_{3})F_{3}J^{3}.
\label{gauge1}%
\end{equation}
Finally the energy over the NCL for the first and third phases is,%

\begin{equation}
E(h,\mu)=\frac{4\pi\Omega}{g}\int_{0}^{\infty}d\xi\left[ \cos^{2}\mu
\frac{v^{2}}{\Omega^{2}}\frac{1}{2}\xi^{2}h^{\prime2}+\xi^{2}\frac{V(h,\mu
)}{g^{2}\Omega^{2}}\right] ,
\end{equation}
and for second phase,
\begin{align}
E(\mu,f,f_{3},f_{0},h) &
=\frac{4\pi\Omega}{g}\int_{0}^{\infty}d\xi\left[
\sin^{2}\mu(\frac{8}{3}{f^{\prime}}^{2}+\frac{4}{3}{f_{3}^{\prime}}^{2}%
)+\frac{8}{\xi^{2}}\sin^{4}\mu\{\frac{2}{3}f_{3}^{2}(1-f)^{2}\right.
\nonumber\\
& +\frac{1}{3}\{f(2-f)-f_{3}\}^{2}\}+\frac{4}{3}(\frac{g}{g^{\prime}}%
)^{2}\{\sin^{2}\mu{f_{0}^{\prime}}^{2}+\frac{2}{\xi^{2}}\sin^{4}\mu
(1-f_{0})^{2}\}\nonumber\\
& +\frac{v^{2}}{\Omega^{2}}\{\frac{1}{2}\xi^{2}{h^{\prime}}^{2}+\frac{4}%
{3}\sin^{2}\mu h^{2}\{(J(J+1)-J_{3}^{2})(1-f)^{2}+J_{3}^{2}(f_{0}-f_{3}%
)^{2}\}\}\nonumber\\
& \left. +\frac{\xi^{2}}{g^{2}\Omega^{4}}V(h)\right] . \label{ensp1}%
\end{align}
From Eq.(\ref{ensp1}), the maximal energy is attained at $\mu=\frac{\pi}{2} $
which corresponds to the sphaleron configuration.

If there are multiple representations $J^{(i)}$ with non-zero neutral
components $J_{3}^{(i)}$,\newline$Q^{(i)}=\frac{v_{i}h_{i}(\xi)}{\sqrt{2}%
}(0,..,1..,0)^{T}$, the energy of the sphaleron can be parameterized as
\begin{align}
E_{sph} & =E(\mu=\frac{\pi}{2})=\frac{4\pi\Omega}{g}\int_{0}^{\infty}%
d\xi\lbrack\left[ \frac{8}{3}{f^{\prime}}^{2}+\frac{4}{3}{f_{3}^{\prime}}%
^{2}+\frac{8}{3\xi^{2}}\{2f_{3}^{2}(1-f)^{2}\right. \nonumber\\
& +(f(2-f)-f_{3})^{2}\}+\frac{4}{3}(\frac{g}{g^{\prime}})^{2}\{{f_{0}%
^{\prime}}^{2}+\frac{2}{\xi^{2}}(1-f_{0})^{2}\}+\sum_{i}\{\frac{1}{2}%
\frac{v_{i}^{2}}{\Omega^{2}}\xi^{2}{h_{i}^{\prime}}^{2}\nonumber\\
& \left. +\frac{4}{3}h_{i}^{2}[2\alpha_{i}(1-f)^{2}+\beta_{i}(f_{0}%
-f_{3})^{2}]\}+\xi^{2}\frac{V(v_{i}h_{i})}{g^{2}\Omega^{4}}\right] ,
\label{ensph}%
\end{align}
where the parameters
\begin{equation}
\alpha_{i}=\frac{(J^{(i)}(J^{(i)}+1)-J_{3}^{(i)2})v_{i}^{2}}{2\Omega^{2}%
},~\beta_{i}=\frac{J_{3}^{(i)2}v_{i}^{2}}{\Omega^{2}}, \label{ab}%
\end{equation}
refer to the scalar field couplings to the charged and neutral gauge fields respectively.

The energy functional, Eq.(\ref{ensph}) will be minimized by the solutions of
the following variational equations%
\begin{gather}
f^{\prime\prime}+\frac{2}{\xi^{2}}(1-f)[f(f-2)+f_{3}(1+f_{3})]+\sum_{i}%
\alpha_{i}h_{i}^{2}(1-f)=0,\nonumber\\
f_{3}^{\prime\prime}-\frac{2}{\xi^{2}}[3f_{3}+f(f-2)(1+2f_{3})]+\sum_{i}%
\beta_{i}h_{i}^{2}(f_{0}-f_{3})=0,\nonumber\\
f_{0}^{\prime\prime}+\frac{2}{\xi^{2}}(1-f_{0})-\frac{{g^{\prime}}^{2}}{g^{2}%
}\sum_{i}\beta_{i}h_{i}^{2}(f_{0}-f_{3})=0,\nonumber\\
h_{i}^{\prime\prime}+\frac{2}{\xi}h_{i}^{\prime}-\frac{8\Omega^{2}}{3v_{i}%
^{2}\xi^{2}}h_{i}[2\alpha_{i}(1-f)^{2}+\beta_{i}(f_{0}-f_{3})^{2}]-\frac
{1}{g^{2}v_{i}\Omega^{2}}\left. \frac{\partial}{\partial\phi_{i}}%
V(\phi)\right\vert _{\phi_{k}=v_{k}h_{k}}=0, \label{Vareq}%
\end{gather}
with the boundary conditions for Eq.(\ref{Vareq}) are given by: $f(0)=f_{3}%
(0)=h(0)=0$, $f_{0}(0)=1$ and $f(\infty)=f_{3}(\infty)=f_{0}(\infty
)=h_{i}(\infty)=1$. For $g^{\prime}\rightarrow0$, we have, $f_{0}%
(\xi)\rightarrow1$ and for representations satisfying Eq.(\ref{rho}),
$f_{3}(\xi)\rightarrow f(\xi)$. The behavior of the field profiles
Eq.(\ref{Vareq}) at the limits $\xi\rightarrow0$ and $\xi\rightarrow\infty
$\ are shown in Appendix \ref{section6}. According to the last term in both
first and second lines in Eq.(\ref{Vareq}), it seems that the couplings of the
scalar to gauge components, i.e. Eq.(\ref{ab}) will play the most important
role in the profile's shape as well as in the sphaleron energy. The equality
between the parameters $\alpha_{i}$ and $\beta_{i}$ leads to the case
Eq.(\ref{rho}) and any difference between $\alpha_{i}$ and $\beta_{i}$ will
characterize a splitting between the functions $f$ and $f_{3}$, and therefore
a departure from the spherical ansatz that was defined in
\cite{Klinkhamer:1984di}.

\subsection{Numerical Results}

\begin{figure}[h]
\begin{centering}
\includegraphics[width=7.5cm,height=6cm]{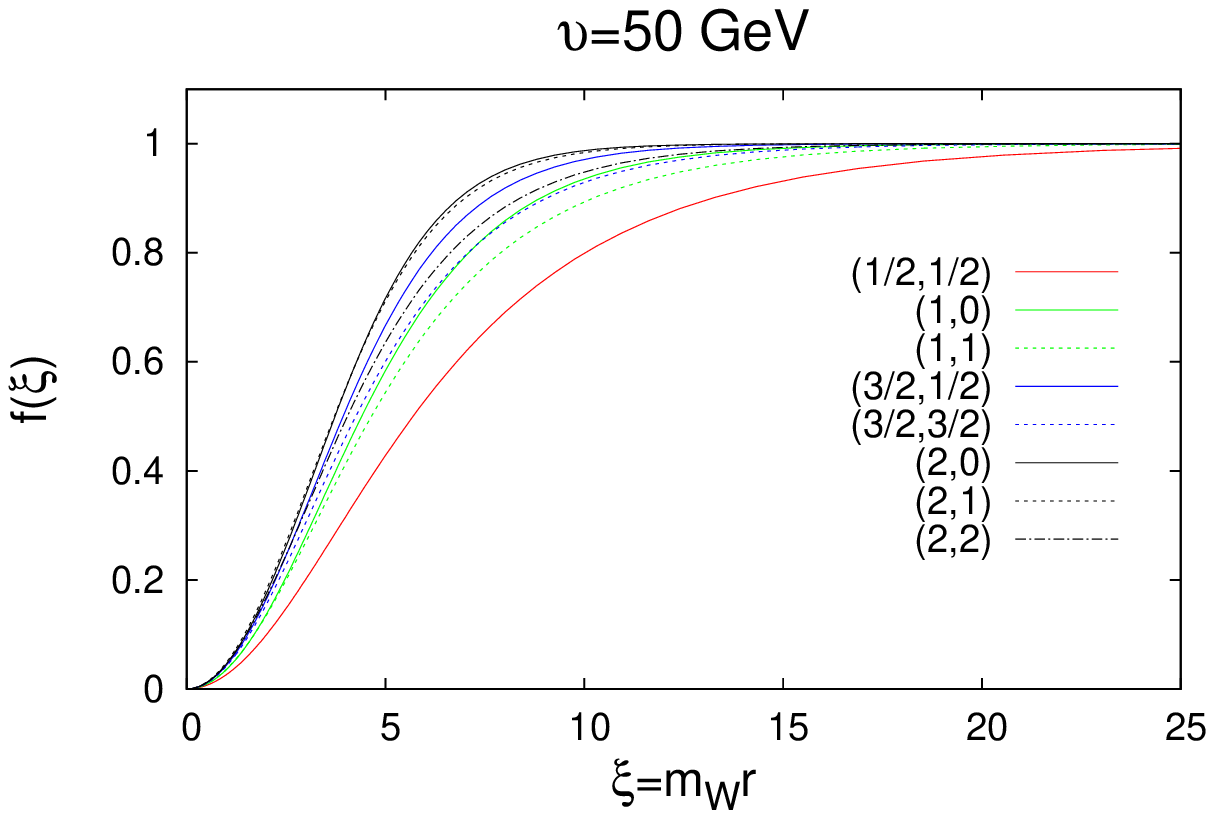}~\includegraphics[width=7.5cm,height=6cm]{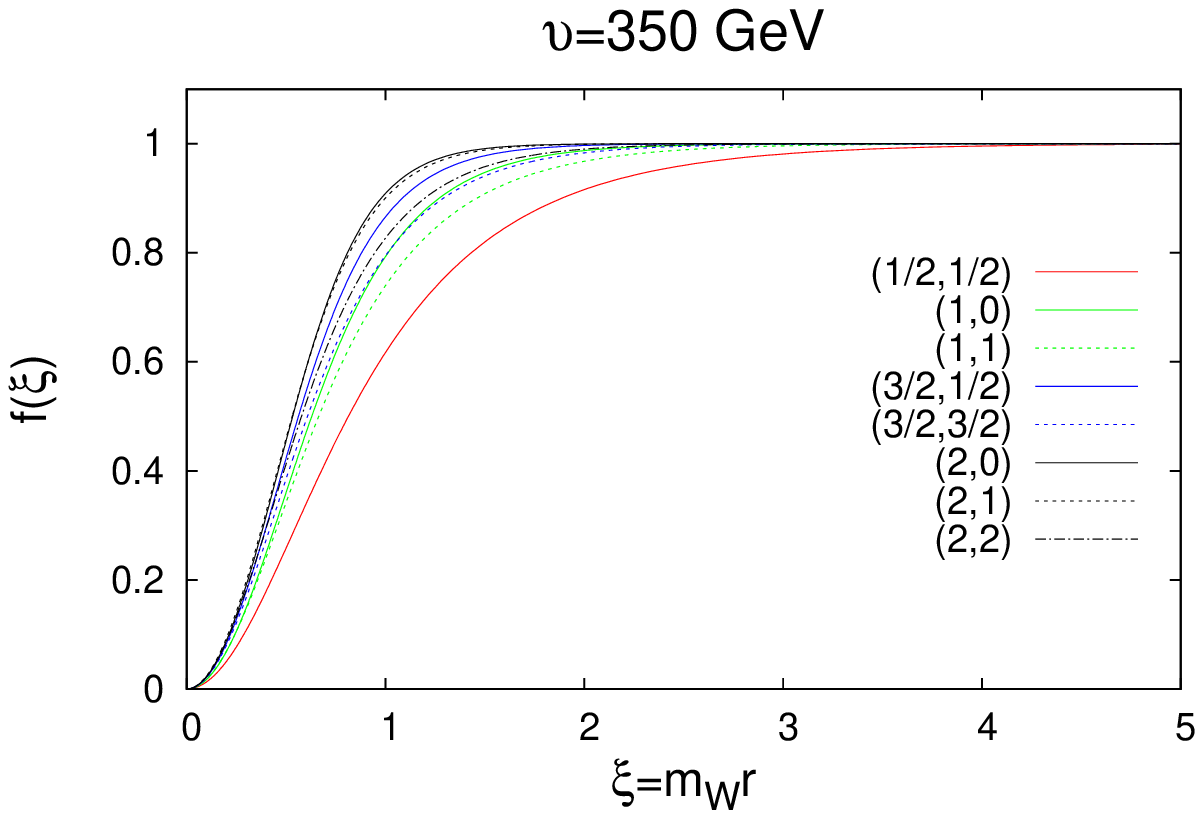}
\includegraphics[width=7.5cm,height=6cm]{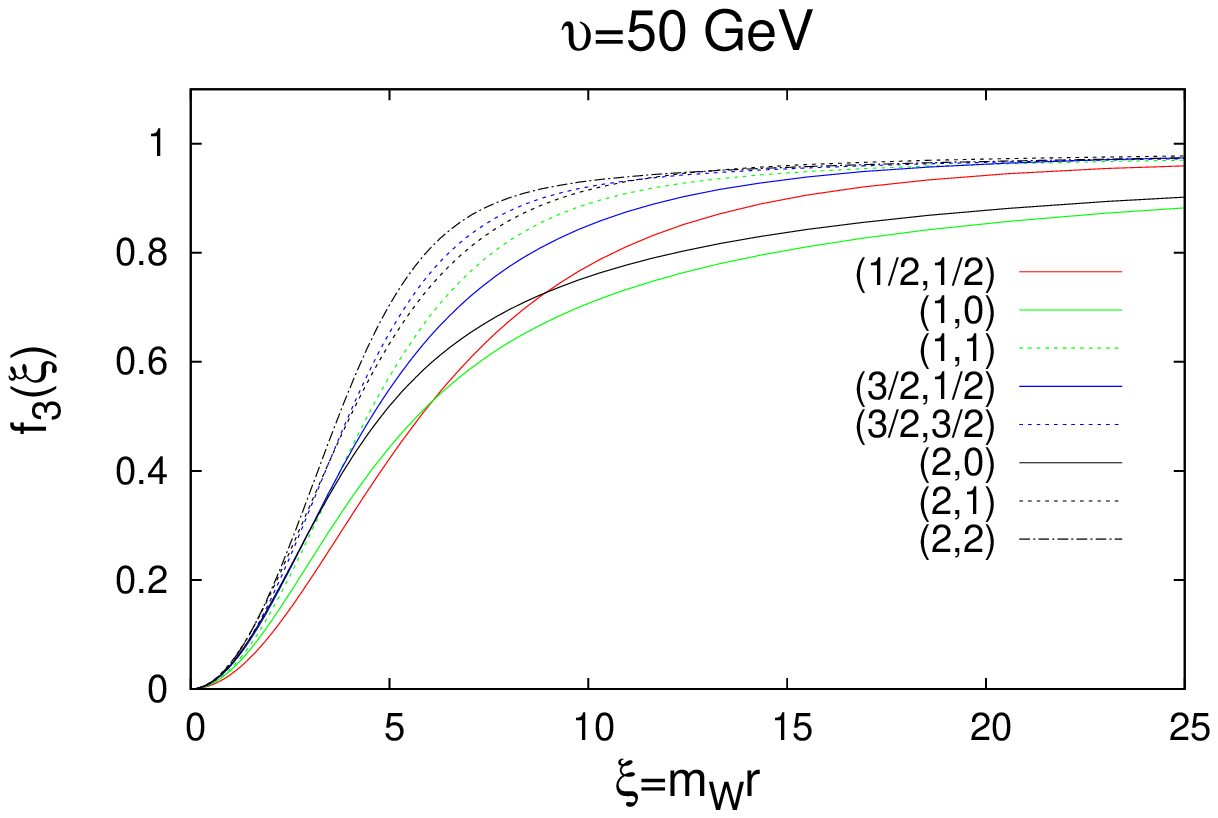}~\includegraphics[width=7.5cm,height=6cm]{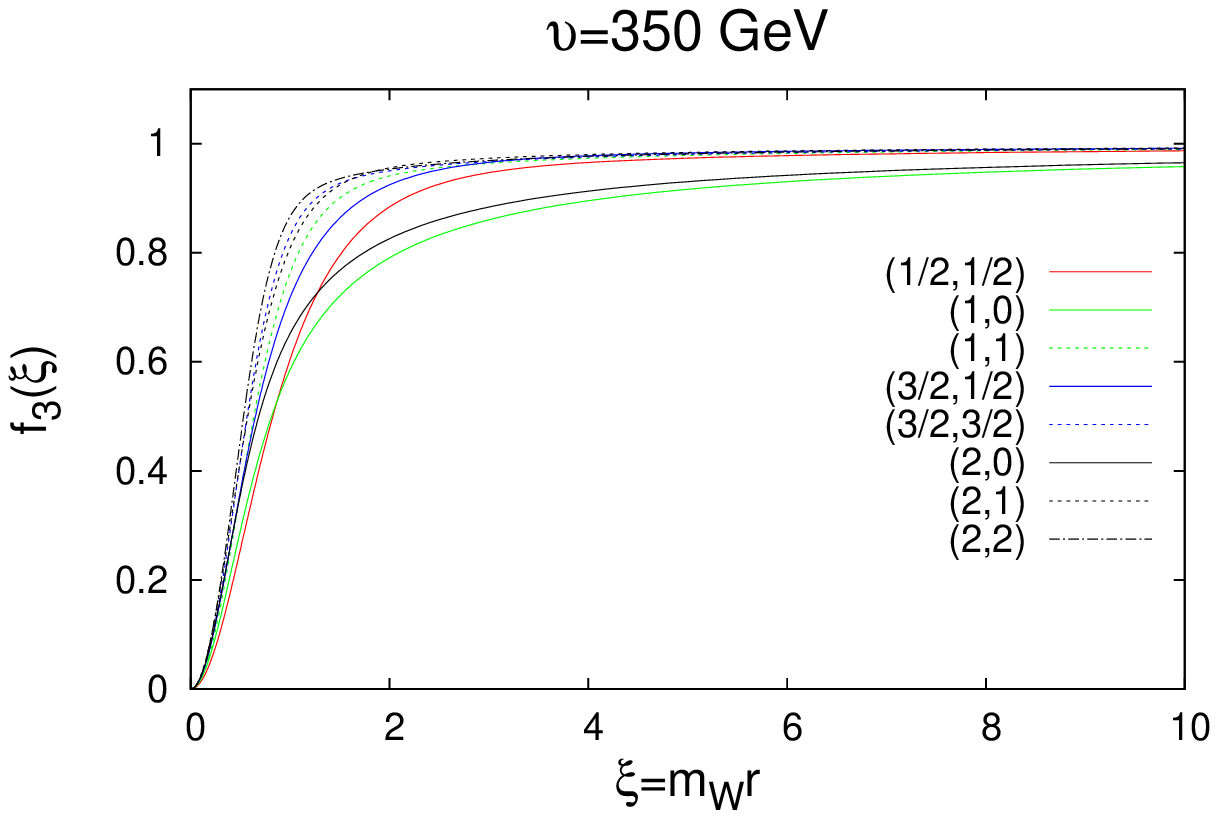}
\includegraphics[width=7.5cm,height=6cm]{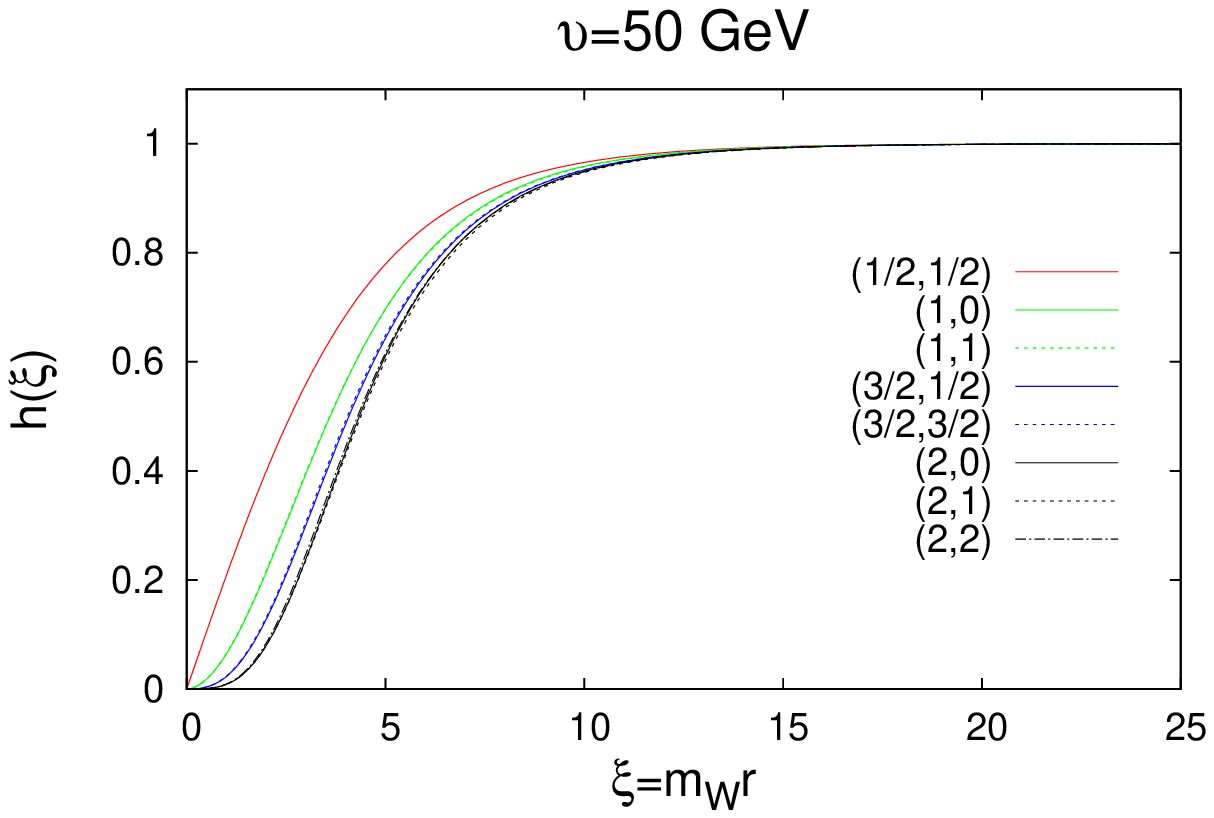}~\includegraphics[width=7.5cm,height=6cm]{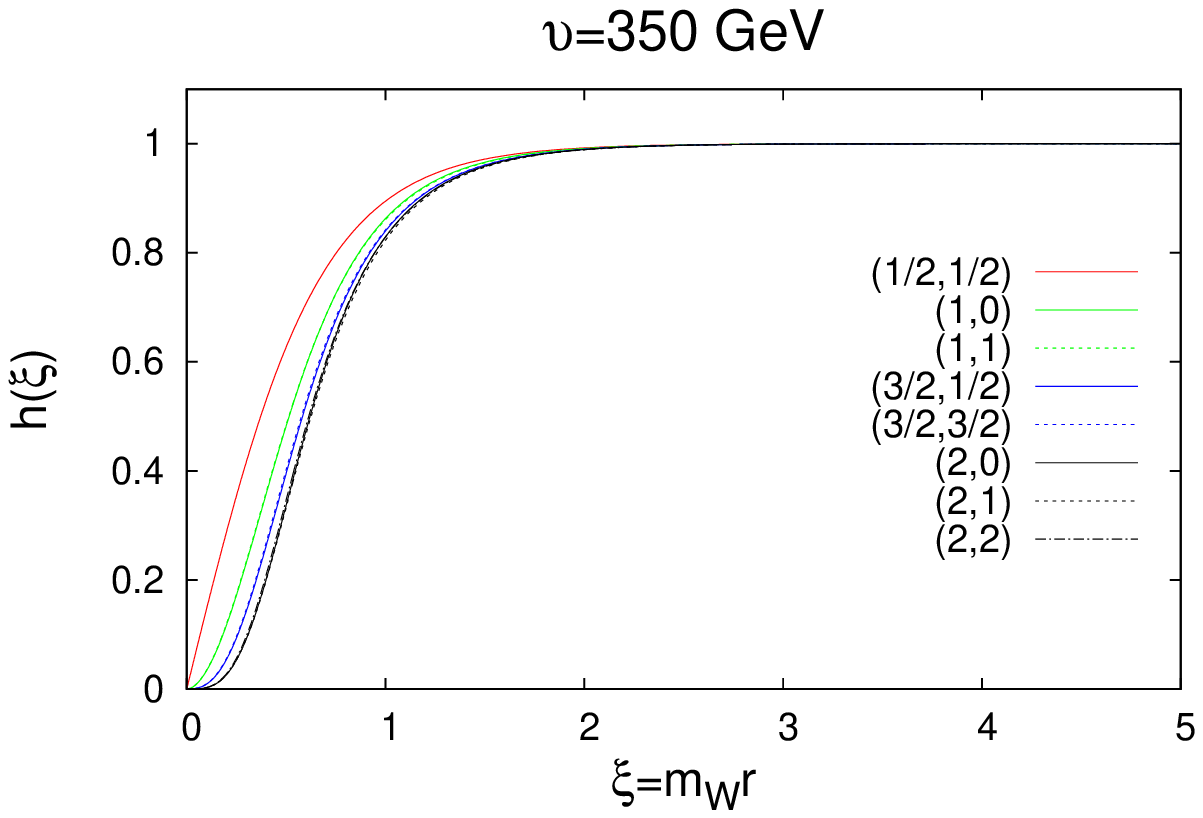}
\par\end{centering}
\caption{The field profiles $f(\xi)$, $f_{3}(\xi)$ and $h(\xi)$ as the
function of the radial coordinate. In the left figures, we set the vacuum
expectation value to be $v=50$ GeV and in the right, it's $v=350$ GeV.}%
\label{Prof}%
\end{figure}

Here we are interested in investigating the properties of the field
profiles for different scalar representations and vevs. First we
have studied the field profiles for only $SU(2)$ with scalar
representation $(J,X)$ where $g^{\prime }$ is taken to be zero and
consequently $f_{0}\rightarrow1$. The scalar representations are
taken as $\left( J,X\right) =\{(1/2,1/2)$, $(1,0)$, $(1,1)$,
$(3/2,1/2)$, $(3/2,3/2)$, $(2,0)$, $(2,1)$, $(2,2)\}$ and two scalar
vevs: $v=50$ \textrm{GeV} and $v=350$ \textrm{GeV}. Here we are
focusing on the sphaleron solution in a generic $SU(2)\times
U(1)_{X}$ case; therefore, we have chosen representative values of
the vev which also contain the SM case, $v=246$ \textrm{GeV} within
the range. Moreover, for each representation, the quartic coupling
is set to be 0.12 and the mass parameter $\mu_{Q}^{2}$ is determined
by coupling and the scalar vev. For this parameter set, the mass of
the scalar field remains smaller than $12m_{W}$ so there is no
appearance of bisphalerons in our case. The field profiles are given
in Figure \ref{Prof}.

According to Figure \ref{Prof}, one can make the following remarks:

\begin{itemize}
\item Comparing the cases of small vev, $v=50$ \textrm{GeV} and large vev,
$v=350$ \textrm{GeV}, it can be seen that all field profiles tend quickly to
the unity as the vev gets larger. This could explain the dependence of
sphaleron energy Eq.(\ref{ensph}) on the scalar vev.

\item When the scalar representation is large (large $J $ so that large
$\alpha$), the profile for charged gauge field (i.e., $f(\xi)$) tends to $1$
faster with $\xi$, in contrast with the scalar field profile, $h(\xi)$.

\item For the neutral gauge field profile $f_{3}(\xi)$, it is identical to
$f(\xi)$ for the representation ($1/2,1/2$) because it satisfies $\rho=1
$\ (or $J(J+1)=3X^{2}$) condition.

\item For the same value of the vev and the isospin $J$, the field profile
$f_{3}(\xi)$ tends to 1 faster for larger values of $J_{3}$, i.e. larger
values of $\beta$.

\item The scalar field profiles $h(\xi)$ seem to be not sensitive to the
values of $J_{3}$.
\end{itemize}

Therefore, it is seen that the gauge field profiles tend to unity faster in
contrast to the scalar field profiles with radial coordinate for large
couplings of the scalar to charged gauge boson, $\alpha$ and neutral gauge
boson, $\beta$. In the next section, we will see the impact of this feature on
the sphaleron energy.

\section{The Effect of $U(1)_{X}$ Field and the Sphaleron Energy}

\label{section3}

In the presence of a non-zero $U(1)_{X}$ gauge coupling $g^{\prime}$ or
non-zero Weinberg angle $\theta_{W}$, the $U(1)_{X}$ gauge field will be
excited and the spherical symmetry will be reduced to axial symmetry. In
\cite{Brihaye:1992jk}, it was shown for the SM with one Higgs doublet that
when the mixing angle is increased, the energy of the sphaleron decreases and
it changes the shape from a sphere at $\theta_{W}=0$ to a very elongated
spheroid at large mixing angle. However, for the physical value of the mixing
angle, the sphaleron differs only little from the spherical sphaleron. On the
other hand, for multiplets not satisfying Eq.(\ref{rho}), the shape of the
corresponding sphaleron will be spheriodal instead of spherically symmetric in
the $SU(2)$ case. In such cases, the large value of the mixing angle may be
significant for the energy and shape of the sphaleron for large multiplet
\cite{future}. In the following, we have adopted the small mixing angle
scenario so that $SU(2)\times U(1)_{X}$ sphalerons are not so different than
the $SU(2)$ case; and we will work at first order of small $\theta_{W}$ value.

\begin{figure}[h]
\begin{centering}
\includegraphics[width=7.5cm,height=6cm]{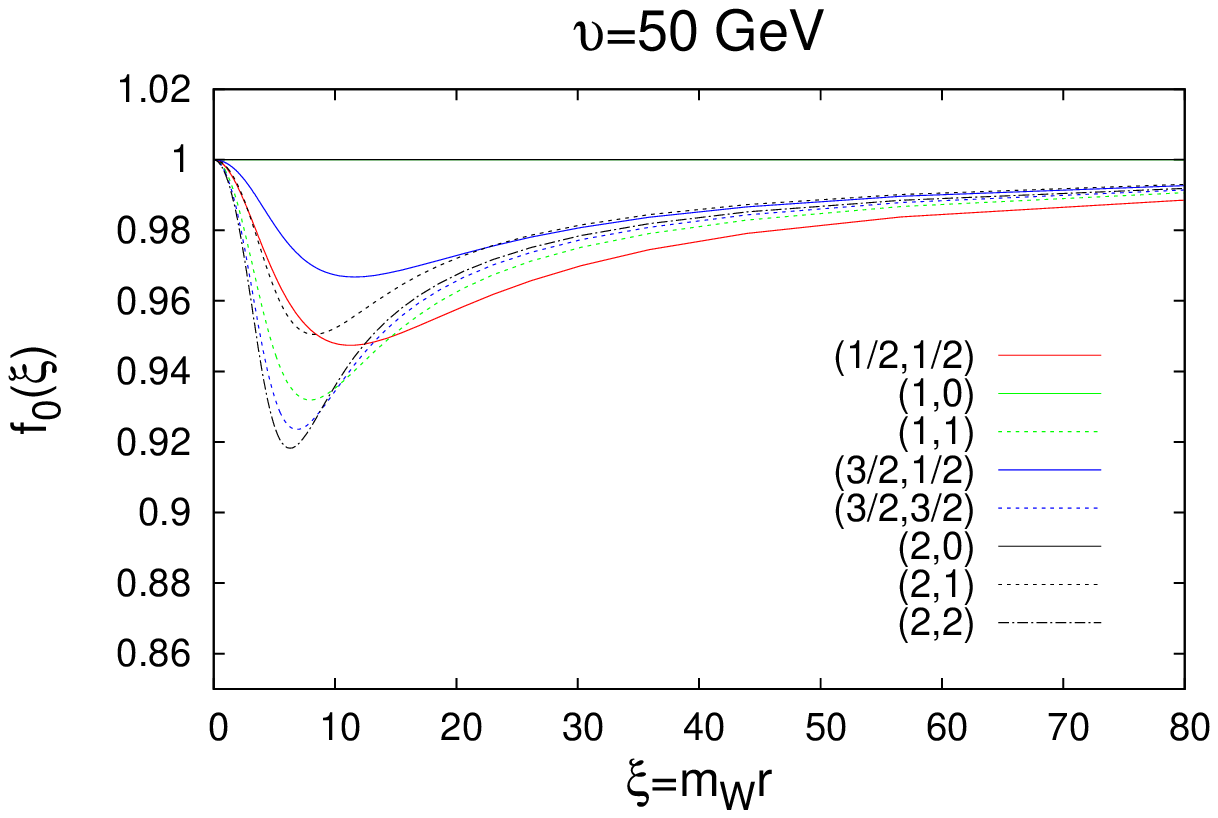}~\includegraphics[width=7.5cm,height=6cm]{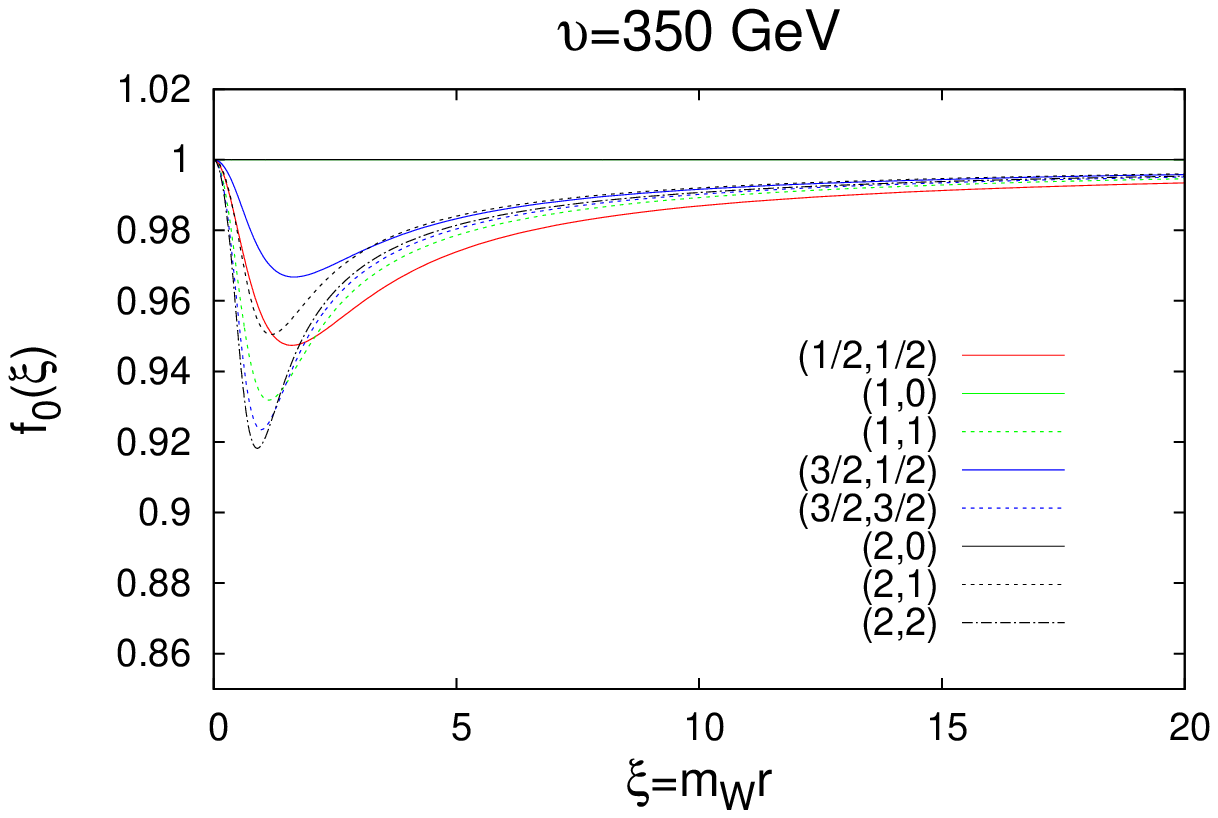}
\par\end{centering}
\caption{The field profile $f_{0}(\xi)$ as a function of the radial
coordinate. In the left figure, we set the vacuum expectation value
to be $v=50$ GeV and in the right, it's $v=350$ GeV.}
\label{f0radial}
\end{figure}

In Figure \ref{f0radial}, we have presented the field profile $f_{0}(\xi)$ for
different values of vev ($v=50,350$ \textrm{GeV}) and different
representations ($J,X$).

In the case of a $SU(2)\times U(1)_{X}$ sphaleron, we have presented only the
field profile $f_{0}(\xi)$ since the other profiles ($f(\xi)$, $f_{3}(\xi)$
and $h(\xi)$) are very close to the case of vanishing Weinberg angle shown in
the previous section. In Figure \ref{f0radial}, one can notice that the field
profile $f_{0}(\xi)$ is just a deviation from unity similar to the singlet
scalar profile in models with singlets \cite{Ahriche:2007jp} and it gets
closer to unity as the $X$ values becomes smaller and smaller. Indeed, it is
exactly one for the representations $(1,0)$ and $(2,0)$ which means that in
those cases the sphaleron energy is not affected by the existence of
$U(1)_{X}$ gauge field.

When we have $\theta_{W}\neq0$, even when one starts with $a_{i}=0$, the
following $U(1)_{X}$ current $j_{i}$ will induce $a_{i}$,
\begin{equation}
j_{i}=\frac{i}{2}g^{\prime}[Q^{\dagger}D_{i}Q-(D_{i}Q)^{\dagger}Q],
\label{u1current}%
\end{equation}

In the leading order approximation of $\theta_{W}$, we can neglect the $a_{i}$
contribution in the covariant derivative. Therefore the non-zero component of
the $U(1)_{X}$ current in the chosen ansatz is \cite{Klinkhamer:1984di}
\begin{equation}
j_{\phi}=\frac{g^{\prime}\sin\theta}{r}\sum_{i}v_{i}^{2}J_{3}^{(i)}h_{i}%
^{2}(1-f). \label{u1current1}
\end{equation}
Because of induced field $a_{i}$, there will be a dipole contribution to the
energy,
\begin{align}
E_{dipole} & =\int d^{3}xa_{i}j_{i}\nonumber\\
& =-\frac{16\pi}{3g\Omega}\sum_{i}v_{i}^{2}J_{3}^{(i)}\int_{0}^{\infty}%
d\xi(1-f_{0})(1-f)h_{i}^{2}, \label{dipole2}%
\end{align}
and the sphaleron energy will be%
\begin{equation}
E_{sph}|_{\theta_{W}\neq0}=E_{sph}|_{\theta_{W}=0}+E_{dipole}.
\end{equation}

\begin{figure}[h]
\begin{centering}
\includegraphics[width=7.5cm,height=5.5cm]{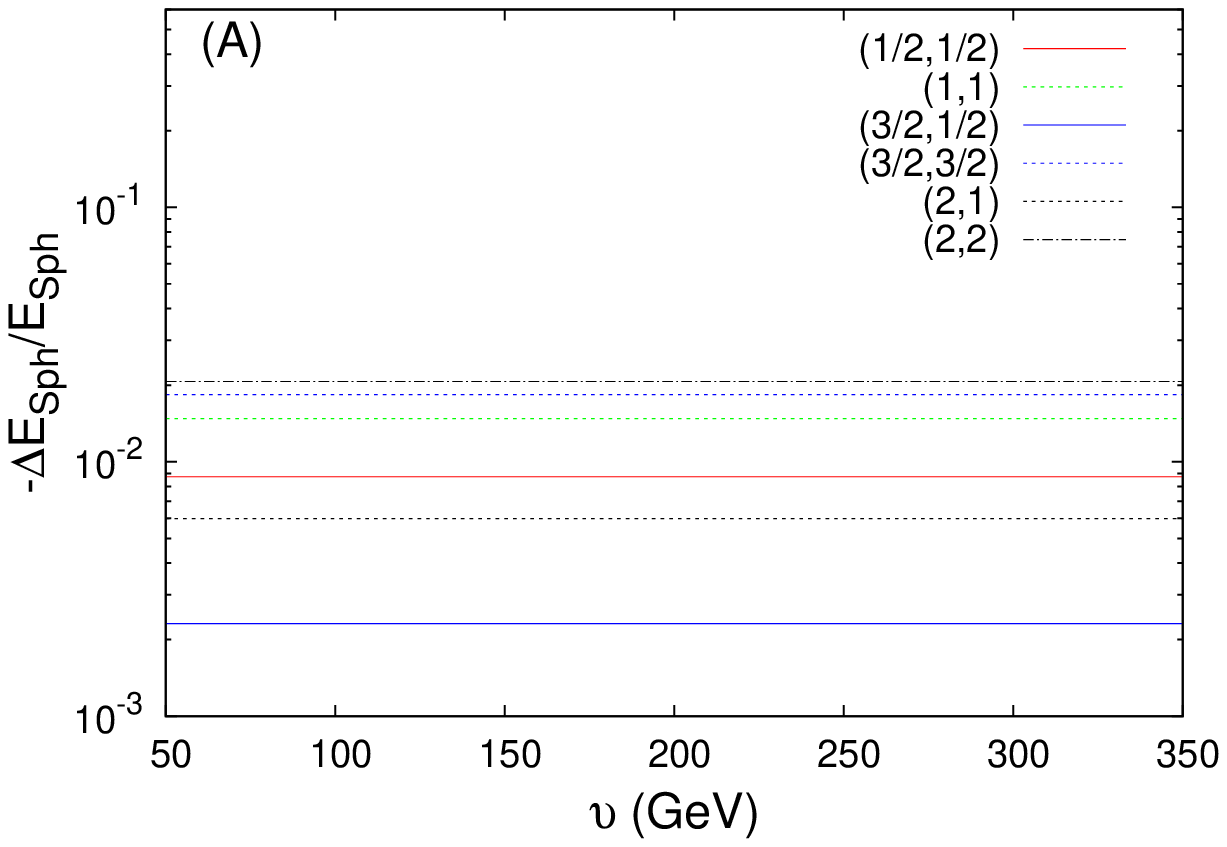}~\includegraphics[width=7.5cm,height=5.5cm]{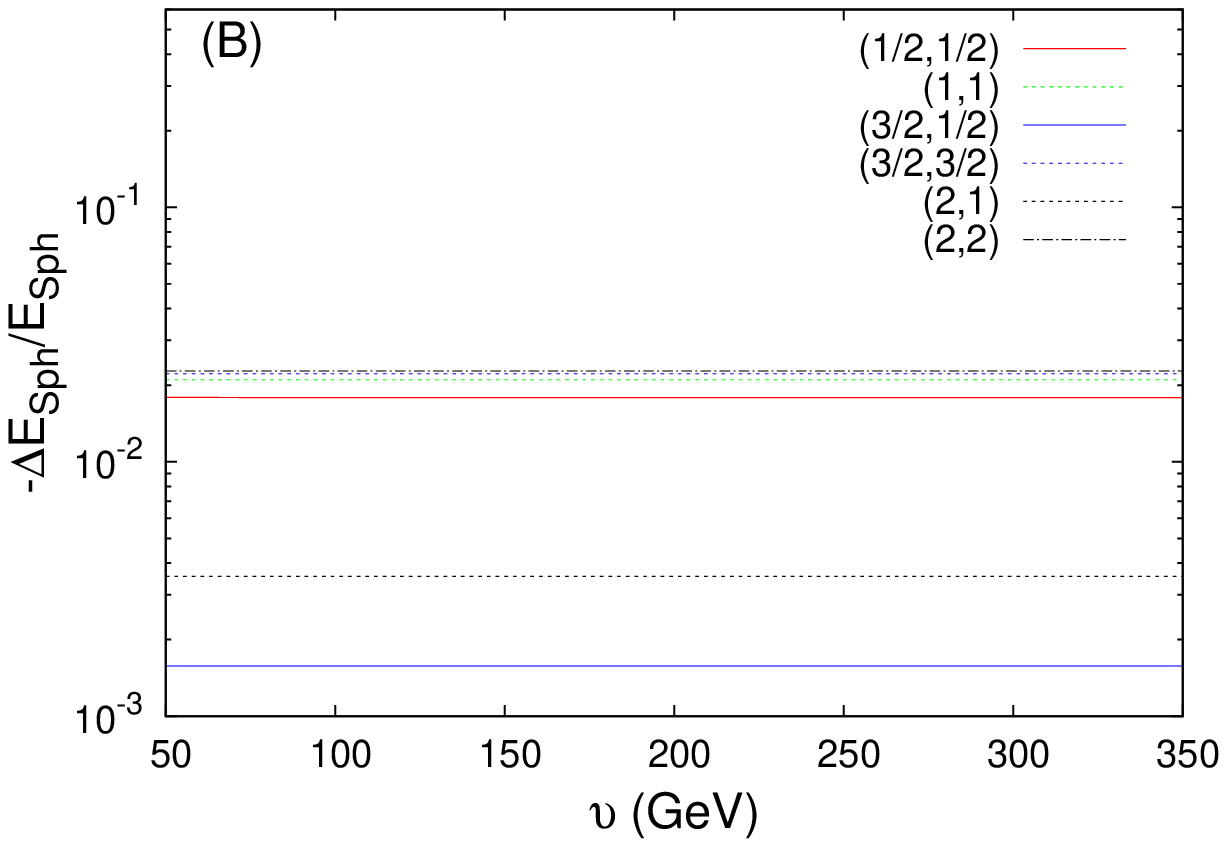}
\includegraphics[width=7.5cm,height=5.5cm]{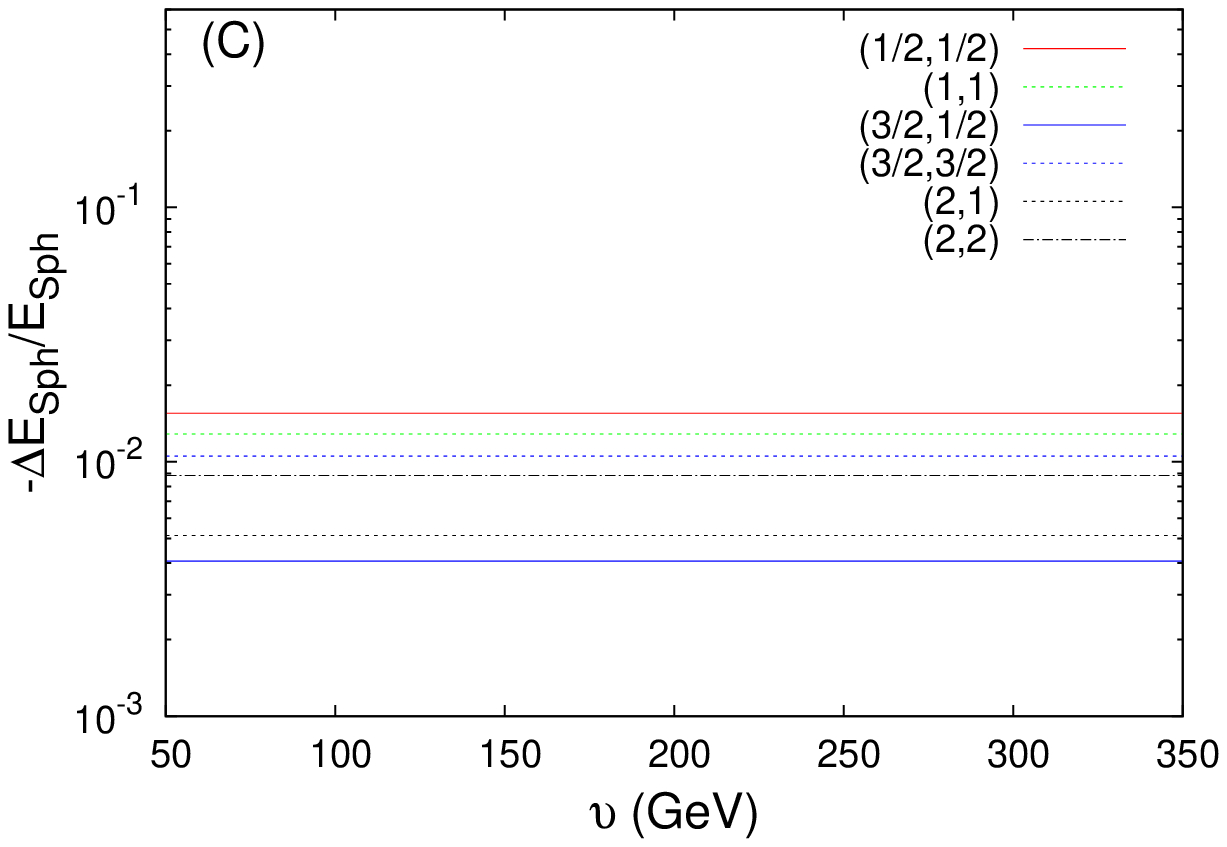}
\par\end{centering}
\caption{The relative difference in the sphaleron energy between the
non-zero and zero mixing cases versus the scalar vev for different
scalar representations, where the difference is estimated: exactly
(left), using the dipole approximation with $U(1)$ gauge field
effect neglected, Eq.(\ref{dipole2}) (right), and the case with
$U(1)$ gauge field effect considered, Eq.(\ref{full}) (down)}
\label{U1}%
\end{figure}

In the current Eq.(\ref{u1current1}) the contribution of the $U(1)_{X}$ gauge
field is generally neglected in the literature and when we consider it, the
current and the dipole energy become
\begin{align}
j_{\phi} & =\frac{g^{\prime}\sin\theta}{r}\sum_{i}v_{i}^{2}J_{3}^{(i)}%
h_{i}^{2}(f_{0}-f_{3}),\nonumber\\
E_{dipole}^{\prime} & =-\frac{16\pi}{3g\Omega}\sum_{i}v_{i}^{2}J_{3}%
^{(i)}\int_{0}^{\infty}d\xi(1-f_{0})(f_{0}-f_{3})h_{i}^{2}, \label{full}%
\end{align}
Therefore the dipole contribution Eq.(\ref{dipole2}) is expected to be almost
equal to the difference between Eq.(\ref{ensph}) and the same quantity with
$g^{\prime}=0$, i.e., $E_{dipole}\simeq\Delta E_{sph}=E_{sph}(g^{\prime}%
\neq0)-E_{sph}(g^{\prime}=0)$. In order to probe this, we estimate the
difference between the sphaleron energy in the non-zero and zero mixing cases
in three different ways: (A) $\Delta E_{sph}=E_{sph}(g^{\prime}\neq
0)-E_{sph}(g^{\prime}=0)$ with $E_{sph}$ is given in Eq.(\ref{ensph}); (B)
$\Delta E_{sph}=E_{dipole}$ with $U(1)_{X}$ field neglected as given in
Eq.(\ref{dipole2}); and (C) $\Delta E_{sph}=E_{dipole}^{\prime}$ as shown in
Eq.(\ref{full}). These three quantities are presented in function of the
scalar vev in Figure \ref{U1}.

Figure \ref{U1} shows the relative difference between the sphaleron energy
with the mixing angle $\theta_{W}\neq0$ and $\theta_{W}=0$ and also the
(negative) dipole energy of the sphaleron. It turns out that for any scalar
representation, the relative difference between the sphaleron energy with
$\theta_{W}\neq0$ and $\theta_{W}=0$ is always less than $1\%$ and remains
constant for different values of scalar vev. However, when considering the
$U(1)_{X}$ gauge field effect on the dipole energy Eq.(\ref{full}), it becomes
closer to the exact difference.

\begin{figure}[h]
\begin{centering}
\includegraphics[width=9cm,height=6cm]{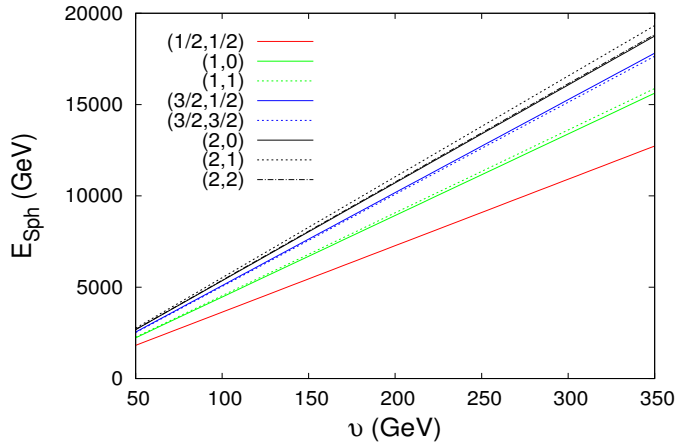}
\par\end{centering}
\caption{The sphaleron energy versus the scalar vev for different
scalar representations.}
\label{EvsV}%
\end{figure}

Now we present the sphaleron energy Eq.(\ref{ensph}) as a function of the
scalar vev for different scalar representations as shown in Figure \ref{EvsV}.

In Figure \ref{EvsV} we can see that the sphaleron energy depends on the
scalar vev with a slope that depends on scalar isospin $J$ and hypercharge $X
$ (or $J_{3}$). This allows us the write the scaling law as
\begin{equation}
E_{sph}(v,J,X)=Z(J,X)\,v, \label{SL}
\end{equation}
where the function $Z(J,X)$ represents the slope in Figure \ref{EvsV}.

\section{Sphaleron Decoupling Condition}

\label{section4}

Before the electroweak phase transition $T>T_{c}$, the classical background
scalar field, $\phi_{c}$, is zero and the Universe is in the symmetric phase.
In this phase, the sphaleron processes \footnote{The term "sphaleron
processes" is used in the literature to refer to the baryon number violating
processes which also have the CP violating feature.} are in full thermal
equilibrium and are given as \cite{Arnold:1996dy, Bodeker:1998hm,
Arnold:1998cy, Moore:2000mx}
\begin{equation}
\Gamma_{sym}\sim\alpha_{w}^{5}T^{4}\ln(1/\alpha_{w}), \label{bv1}%
\end{equation}
with $\alpha_{w}=g^{2}/4\pi$ is the weak coupling. Therefore any
generated baryon asymmetry due to the sphaleron processes will be
erased by the inverse process. Once the temperature drops below the
critical one $T<T_{c}$, bubbles of true vacuum ($\phi_{c}\neq0$)
start to nucleate where the rate is suppressed as
$\Gamma\sim\exp\left( -E_{sph}/T\right) $.

The sphaleron decoupling condition indicates that the rate of baryon
number violation must be much smaller than the the Hubble parameter
\cite{Shaposhnikov:1987tw, Shaposhnikov:1987pf, Bochkarev:1987wf}
and therefore, the condition on the sphaleron rate is
\cite{Arnold:1987mh, Akiba:1989xu, Funakubo:2009eg}
\begin{equation}
-\frac{1}{B}\frac{dB}{dt}\simeq\frac{13N_{f}}{128\pi^{2}}\frac{\omega_{-}%
}{\alpha_{w}^{3}}\kappa{\mathcal{N}_{tr}}{\mathcal{N}_{rot}}e^{-E_{sph}%
/T}<H(T), \label{sphaleron1}%
\end{equation}
where $B$ is the baryon number density, the factors $N_{tr}~$and $N_{rot}$
come from the zero mode normalization, $\omega_{-}$ is the eigenvalue of the
negative mode \cite{Carson:1989rf}. The factor $\kappa$ is the functional
determinant associated with fluctuations around the
sphaleron~\cite{Dine:1991ck}. It has been estimated to be in the range:
$10^{-4}\lesssim\kappa\lesssim10^{-1}$~\cite{Carson:1990jm,kapan}. The Hubble
parameter is given as
\begin{equation}
H(T)\simeq1.66\sqrt{g_{\ast}(T)}T^{2}/M_{pl}, \label{sphaleron2}%
\end{equation}
where $M_{pl}$ and $g_{\ast}$ are the Planck mass and the effective number of
degrees of freedom that are in thermal equilibrium.

It was shown in \cite{Braibant:1993is} for the doublet case $(J,X)=(1/2,1/2)$
that the sphaleron energy at a given temperature can be well approximated by
the following relation
\begin{equation}
\frac{E_{sph}(v(T),T)}{v(T)}=\frac{E_{sph}(v_{0})}{v_{0}}, \label{sphscale}%
\end{equation}
where $v(T)$ is the vev of the scalar field at temperature $T$ and $v_{0}$ is
its zero temperature value. Eq.(\ref{sphscale}) shows that a straightforward
estimation of the sphaleron energy at finite temperature is possible by
determining its energy at zero temperature. This means that the scaling law
Eq.(\ref{SL}) is valid also at finite temperature case, where the function
$Z(J,X)$ is temperature-independent. Because of similar linear scaling shown
by higher scalar representations in Figure \ref{EvsV}, we can use the scaling
law Eq.(\ref{SL}) for other representations.

Hence, for general scalar representation, the decoupling of baryon number
violation Eq.(\ref{sphaleron1}) implies the following relation
\cite{Funakubo:2009eg}
\begin{equation}
\frac{v(T_{c})}{T_{c}}>\frac{1}{Z(J,X)}\left[ 42.97+\ln(\kappa{\mathcal{N}%
_{tr}}{\mathcal{N}_{rot}})+\ln\tfrac{\omega_{-}}{m_{W}}-\tfrac{1}{2}\ln
\tfrac{g_{\ast}}{106.75}-2\ln\tfrac{T_{c}}{100\text{ {\normalsize \textrm{GeV
}}}}\right] . \label{sphaleron4}%
\end{equation}
Most of the parameters in the r.h.s of Eq.(\ref{sphaleron4}) are
logarithmically model-dependent and therefore one can safely use the SM
values. In the case of SM, we have $\mathcal{N}_{tr}\mathcal{N}_{rot}%
\simeq80.13 $ \cite{Arnold:1987mh} and for $\lambda/g^{2}=1$, $\omega_{-}%
^{2}\simeq2.3m_{W}^{2}$ \cite{Carson:1989rf, Carson:1990jm, Akiba:1989xu}. It
can be noted that the contributions of model dependent quantities in $v(T)/T$
are smaller than $Z(J,X)$, for example, in the SM \cite{Funakubo:2009eg} zero
mode contribution is around $10\%$ and the contributions from the negative
mode, relativistic degrees of freedom and critical temperature are about
$1\%$. For this reason we can consider the dominant contribution is coming
from $Z(J,X)$. In conjunction, using $\kappa=10^{-1}$ (or $10^{-4}$),
$g_{\ast}\simeq106.75$ and $T_{c}\simeq100$ \textrm{GeV}, we have from
Eq.(\ref{sphaleron4}),%
\begin{equation}
\frac{v(T_{c})}{T_{c}}>\eta_{J,X}, \label{eta}%
\end{equation}
where $\eta_{J,X}$ is given for each scalar representation in Table-\ref{T3}.

\begin{table}[tbh]
\begin{center}%
\begin{tabular}
[c]{cccccc}\hline\hline $J$ & $X$ & $Z(J,X)$ & $\eta_{J,X}\left(
\kappa=10^{-4}\right) $ & & $\eta_{J,X}\left( \kappa=10^{-1}\right)
$\\\hline 1/2 & 1/2 & 36.37 & $\allowbreak1.0601$ & &
$1.2500$\\\hline
1 & 0 & 44.64 & $0.8639$ & & $1.0186$\\
& 1 & 45.37 & $\allowbreak0.8500$ & & $1.0023$\\\hline
3/2 & 1/2 & 50.89 & $\allowbreak0.7577$ & & $0.8934$\\
& 3/2 & 50.42 & $\allowbreak0.7648$ & & $\allowbreak0.9018$\\\hline
2 & 0 & 53.58 & $0.7197$ & & $0.8486$\\
& 1 & 55.22 & $\allowbreak0.6984$ & & $\allowbreak0.8235$\\
& 2 & 53.80 & $\allowbreak0.7167$ & &
$\allowbreak0.8451$\\\hline\hline
\end{tabular}
\end{center}
\caption{The values for the parameters $Z(J,X)$ and $\eta_{J,X}$ for different
scalar representations.}%
\label{T3}%
\end{table}

\begin{figure}[h]
\begin{centering}
\includegraphics[width=9cm,height=7cm]{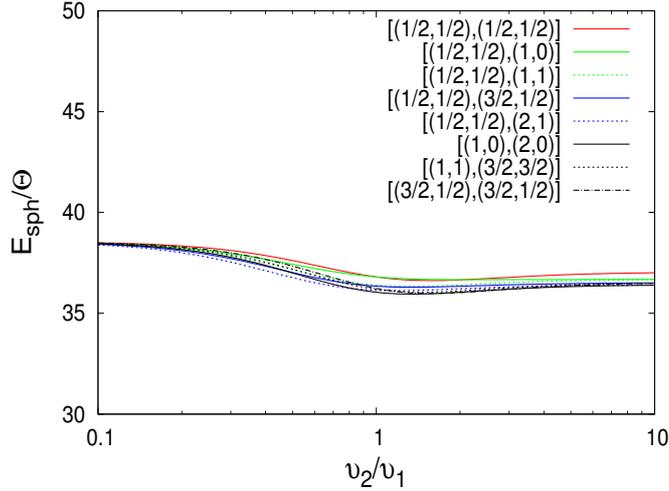}
\par\end{centering}
\caption{The sphaleron energy versus the scalar vev for different
scalar representations. The self quartic couplings of scalar
multiplet $Q_{1}(J_{1},X_{1})$ ($Q_{2}(J_{2},X_{2})$) is set to
$0.12$ ($0.06$) while the mixing quartic coupling is set to $0.02$.}
\label{Tt}
\end{figure}

It is clear that as the representation becomes larger, the strong first order
phase transition criterion gets relaxed. Generally, the case of $\kappa
=10^{-4} $ is the commonly used criterion in the literature. In a general case
of a multi-scalars model with representations ($J^{(i)},X^{(i)} $), the
criterion Eq.(\ref{eta}) can be generalized as%
\begin{equation}
\frac{\Theta(T_{c})}{T_{c}}>1, \label{Nc}%
\end{equation}
with%
\begin{equation}
\Theta(T_{c})^{2}=\sum\limits_{i}\frac{v_{i}(T_{c})^{2}}{\eta_{J^{(i)}%
,X^{(i)}}^{2}}, \label{Theta}%
\end{equation}
with $v_{i}(T)$ is the temperature dependent scalar vev of the multiplet
$Q^{(i)}$. In order to check the criterion Eq.(\ref{Theta}), we consider the
case of a model with two scalar representations and estimate the ratio
$E_{sph}/\Theta$ for different values of $J_{1}$, $J_{2}$, $X_{1}$, $X_{2}$,
$v_{1}$ and $v_{2}$ while keep the W gauge boson mass constant. The ratio
$E_{sph}/\Theta$ versus the ratio $v_{2}/v_{1}$ is shown in Figure \ref{Tt}.

From Figure \ref{Tt}, it is clear that the sphaleron energy scales like
$\Theta$ for different representations and vevs within the error less than 5.7
\%; and if the values of the two vevs are comparable, this error is reduced to
2.7 \%. Therefore, one can safely use Eq.(\ref{Theta}) as a criterion for a
strong first order phase transition in any model with multiscalars.

\section{Conclusion}

\label{section5}

We have constructed the energy functional and relevant variational equations
of the sphaleron for general scalar representation charged under $SU(2)\times
U(1)_{X}$ gauge group and shown that the sphaleron energy increases with the
size of the multiplet. Furthermore, it has been shown that at a fixed value of
the vev, the sphaleron energy is large for larger representation and for each
representation, it linearly scales with the vev. As the energy of the
sphaleron increases with the size of the scalar representation, the criterion
for the strong first order phase transition is relaxed for larger
representation. We have presented a representation dependent criterion for
strong phase transition which is relevant for the electroweak baryogenesis.

We have also found that the dipole approximation (with or without considering
$a_{i}$ in the $U(1)_{X}$ current, $j_{i}$) does not correspond exactly the
energy difference $E_{sph}(g^{\prime}\neq0)-E_{sph}(g^{\prime}=0)$ and that is
less than 2\% for any scalar representation. In this case the $U(1)_{X}$ field
profile is just a deviation from unity and therefore just playing a relaxing
role similar to singlet seen in \cite{Ahriche:2007jp}.

However, as we have seen in Figure \ref{U1} that the dipole contribution to
the sphaleron energy is negative, its coupling with the external magnetic
field produced in the bubbles of first order phase transition through the
dipole moment would lower the sphaleron energy and thus strengthen the
sphaleron transition inside the bubble and wash out the baryon asymmetry more
efficiently as pointed out in \cite{DeSimone:2011ek}. A more careful analysis
on this aspect for the sphaleron with higher scalar representation will be
carried out in \cite{future}.

We have presented in Eq.(\ref{Theta}) a general criterion for the strong first
order phase transition in a model with multiple scalars of different
representations $(J,X)$ and we have shown that this approximate criterion is
valid with an error less than 5\%.

\section*{Acknowledgements}

We are grateful to Goran Senjanovi\'c, Eibun Senaha, Andrea De Simone,
Dietrich B\"{o}deker and Xiaoyong Chu for the critical reading of the
manuscript and helpful comments. T.A.C. would also like to thank Basudeb
Dasgupta, Luca Di Luzio and Marco Nardecchia for discussion. A.A. is supported
by the Algerian Ministry of Higher Education and Scientific Research under the
CNEPRU Project No. \textit{D01720130042}.

\appendix

\section{Asymptotic solutions}

\label{section6}

To capture the dependence of solutions on $(J,X)$, in this section we have
included the analytical estimates of solutions for the asymptotic region
$\xi\rightarrow0$ and $\xi\rightarrow\infty$. For the energy functional
Eq.(\ref{ensph}) to be finite, the profile functions should be $f(\xi
)\rightarrow0$, $f_{3}(\xi)\rightarrow0$, $f_{0}(\xi)\rightarrow1$ and
$h(\xi)\rightarrow0$. Therefore, at $\xi\sim0 $, the equations Eq.(\ref{Vareq}%
) are reduced into
\begin{align}
\xi^{2}f^{\prime\prime} & -4f+2f_{3}+\alpha\xi^{2}h^{2}=0,\label{eq1}\\
\xi^{2}f_{3}^{\prime\prime} & -6f_{3}+4f+\beta\xi^{2}h^{2}=0,\label{eq2}\\
f_{0}^{\prime\prime} &
+2(1-f_{0})-(\frac{g^{\prime}}{g})^{2}\beta\xi
^{2}h^{2}=0,\label{eq3}\\
\xi^{2}h^{\prime\prime} & +2\xi h^{\prime}-\frac{8m}{3}h=0, \label{eq4}%
\end{align}
where
\begin{equation}
m=\frac{\Omega^{2}}{v^{2}}(2\alpha+\beta). \label{def1}%
\end{equation}
The solution of Eq.(\ref{eq4}) which leads to finite energy of the sphaleron
is
\begin{equation}
h(\xi)\sim A\xi^{-\frac{1}{2}(1-p)}, \label{eq5}%
\end{equation}
with
\begin{equation}
p=\sqrt{1+\frac{32}{3}m}. \label{def2}%
\end{equation}
Now at $\xi\sim0$, $f(\xi)\sim f_{3}(\xi)$, so using this approximation, from
Eq.(\ref{eq1}) we have,%

\begin{equation}
f(\xi)\sim B\xi^{2}-\frac{4A\alpha\xi^{\frac{1}{2}(3+p)}}{(\frac{p}%
{2}-1)(\frac{p}{2}+5)}. \label{eq6}%
\end{equation}
On the other hand, we have considered $f(\xi)$ as a perturbation in
Eq.(\ref{eq2}). Therefore, we have
\begin{equation}
f_{3}(\xi)\sim C\xi^{3}+B\xi^{2}-K\xi^{\frac{1}{2}(3+p)}. \label{eq7}%
\end{equation}
Here, $K$ is defined as follows%
\begin{equation}
K=\frac{3A\{3\alpha(3p-8m+3)+8m\beta(4m-9)\}}{4m(4m-9)(8m+3p-15)}.
\end{equation}
Finally from Eq.(\ref{eq3}), we have
\begin{equation}
f_{0}(\xi)\sim1+D\xi^{2}+\frac{3A\beta g^{\prime2}\xi^{\frac{1}{2}(3+p)}%
}{g^{2}(3p-8m+3)}, \label{eq8}%
\end{equation}
and $A$, $B$, $C$ and $D$ are integration constants.

On the other hand, for asymptotic region, $\xi\sim\infty$, all the profile
functions must approach unity to have finite energy of the sphaleron. So we
consider the functions to be the small perturbation to unity as follows.
Taking, $f(\xi)=1+\delta f(\xi)$, $f_{3}(\xi)=1+\delta f_{3}(\xi)$, $f_{0}%
(\xi)=1+\delta f_{0}(\xi)$ and $h(\xi)=1+\delta h(\xi)$ and keeping only the
linear terms of the variation, we have%
\begin{align}
\delta f^{\prime\prime}-\alpha\delta f & =0,\nonumber\\
\delta f_{3}^{\prime\prime}+\beta(\delta f_{0}-\delta f_{3}) &
=0,\nonumber\\
\delta f_{0}^{\prime\prime}-\frac{g^{\prime2}}{g^{2}}\beta(\delta f_{0}-\delta
f_{3}) & =0,\nonumber\\
\xi^{2}\delta h^{\prime\prime}-2\xi\delta h-3\frac{\lambda v^{2}}{g^{2}%
\Omega^{2}}\xi^{2}\delta h & =0.
\end{align}
The asymptotic solutions at $\xi\sim\infty$ are,
\begin{align}
f(\xi) & \sim1+Ee^{-\sqrt{\alpha}\xi},\nonumber\\
f_{3}(\xi) & \sim1+Fe^{-\sqrt{\beta}\xi},\nonumber\\
f_{0}(\xi) & \sim1+Ge^{-\sqrt{\beta}\xi},\nonumber\\
h(\xi) &
\sim1+\frac{He^{-\frac{\sqrt{3\lambda}v}{g\Omega}\xi}}{\xi},
\end{align}
where $E$, $F$, $G$ and $H$ are again integration constants. The constants
from $A$ to $H$ depend on $(J,X)$ and couplings and they are determined by
matching the corresponding asymptotic solutions and their first derivatives at
$\xi=0$. Therefore after the matching, the integration constants are,$v_{1}$
and $v_{2}$%

\begin{align}
H &
=-\frac{\frac{1}{2}(p-1)e^{\frac{v}{\Omega}n}}{\frac{1}{2}(p+1)+\frac
{v}{\Omega}n},~A=1+He^{-\frac{v}{\Omega}n},\nonumber\label{const1}\\
E & =-\frac{e^{\sqrt{\alpha}}}{\sqrt{\alpha}+2}(2+\frac{2A\alpha
(1-p)}{(\frac{p}{2}-1)(\frac{p}{2}+5)}),~B=1+Ee^{-\sqrt{\alpha}}%
+\frac{4A\alpha}{(\frac{p}{2}-1)(\frac{p}{2}+5)},\nonumber\\
B & =1+Ee^{-\sqrt{\alpha}}+\frac{4A\alpha}{(\frac{p}{2}-1)(\frac{p}{2}%
+5)},~F=\frac{e^{\sqrt{\beta}}}{\sqrt{\beta}+3}(-3+B-\frac{1}{2}%
(3-p)K),\nonumber\\
C & =1+Fe^{-\sqrt{\beta}}-B+K,~G=\frac{e^{\sqrt{\beta}}}{\sqrt{\beta}%
+2}\frac{3A\beta n_{1}^{2}(1-p)}{2(3p+8m-3)},\nonumber\\
D & =Ge^{-\sqrt{\beta}}-\frac{3A\beta n_{1}^{2}}{3p+8m-3},
\end{align}
where $n=\sqrt{3\lambda}/g$, where $\lambda$\ is the scalar quartic coupling.

\end{document}